\documentstyle[aps,amssymb,twocolumn]{revtex}

\begin{document}
\title{Modulation origin of $1/f$ noise in two-dimensional hopping}
\author{V.Pokrovskii$^{1,2}$, A.K. Savchenko$^1$, W.R.Tribe$^3$, E.H.Linfield$^3$}
\address{$^{1}$ School of Physics, University of Exeter,  Exeter, EX4 4QL, UK\\
$^2$ Inistitute of Radio Engineering and Electronics, 103907\\
Moscow, Russia\\
$^3$ Cavendish Laboratory, University of Cambridge, Cambridge CB3
0HE, UK} \maketitle

\begin{abstract}
We show that $1/f$ noise in a two-dimensional electron gas with
hopping conduction can be explained by the modulation of
conducting paths by fluctuating occupancy of non-conducting
states. The noise is sensitive to the structure of the critical
hopping network, which is varied by changing electron
concentration, sample size and temperature. With increasing
temperature, it clearly reveals the crossover between different
hopping regimes.
\end{abstract}

\vskip 1cm

The problem of $1/f$ noise in hopping conduction has been
attracting much attention for nearly 20 years, with several models
proposed to explain the conductance fluctuations
\cite{KoShk,Kozub,Kogan}. In the single-electron approach the
hopping conduction is determined by the critical network
(percolation cluster) connecting the pairs of localised states
\cite{SE}. In the earlier model \cite{KoShk}, developed for
nearest-neighbour hopping (NNH), it was suggested that $1/f$ noise
originates from fluctuations in the electron concentration on the
cluster, caused by slow electron exchange with the states in its
pores. This theory predicts a saturation in the noise spectrum at
low frequencies, which has not been seen so far in experiment. In
an alternative approach \cite{Kozub}, resistance fluctuations come
from the change in the hopping paths themselves, due to
fluctuations of the energy levels of the localised states in the
cluster. These fluctuations are caused by the random electric
field generated by the states outside the cluster, due to the
random occupancy of the states. To explain the absence of the
saturation predicted in \cite{KoShk}, a configuration model was
proposed \cite{Kogan}. In this model $1/f$ noise is treated as a
many-electron problem and is explained by transitions between
metastable states of the interacting hopping system.

On the experimental side, it was shown on Si inversion layers in
\cite{Voss} that $1/f$ noise is an intrinsic property of hopping
conduction. In experiments on mesoscopic {\em n} -GaAs samples
\cite{AKS}, an indication was obtained that the energy level
fluctuations are important in the origin of hopping noise. Large
random telegraph signals (RTSs) were observed, and it was
suggested that they are produced by the random field due to
low-frequency, long-distance hops between states outside the
cluster. The modulation theory of noise \cite{Kozub} has
considered the integral effect of all modulators in large samples.
Here, the temperature dependence of the resistance noise
$S_R/R^{2}(T)$ has been derived, although the experiments
performed on 3D samples \cite {Shlimak,MarkLee} did not show a
unanimous agreement with the theory. While a weak temperature
dependence of $1/f$ noise in the NNH regime \cite{Shlimak} was
used in \cite{Kozub} as support for the modulation theory, the
measurements of noise in variable-range hopping (VRH) \cite
{MarkLee} have shown a temperature dependence of opposite sign to
that predicted in \cite{Kozub}, and were interpreted as a
consequence of electron-electron interaction.

In this work we study $1/f$ noise in 2D hopping of an electron
channel in a MESFET structure, in a broad range of temperatures,
from 4.2 to 60 K. In this versatile transistor system, we perform
experiments at different electron concentration, varied by the
gate voltage. This allows us to vary the resistance and topology
of the critical cluster in the same sample, and see how these
changes affect the noise. Additional control of the conducting
network is achieved by changing the sample length: in the shortest
sample with $L\simeq 0.2$ $\mu $m the cluster is reduced down to a
set of well separated linear chains \cite{AKSmeso}. Our results
show that in the VRH regime the character of the cluster is
reflected in $1/f$ noise, and it can be well described in terms of
the single-electron modulation approach. Also, the noise
properties appear to be very sensitive to the transition between
different hopping regimes, observed when the temperature is
increased.

We have studied three samples, with lengths $L=$ 0.2, 0.5 and
1~$\mu $m and width $W=$ 5~$\mu $m arranged as individual
transistors on a uniformly doped, 1500 \AA\ thick GaAs layer with
donor concentration $\simeq 10^{17}cm^{-3}$. At zero gate voltage
{\bf $V_g$,} the channel has metallic conduction, and with its
depletion a transition to strong localisation occurs
\cite{Pepper}. The thickness $t$ of the channel in the depleted
regime is close to the mean donor separation ($\approx $ 200~\AA )
and satisfies the condition of 2D hopping: $t\leq r$, where $r$ is
the typical hopping distance. The resistance noise is measured as
fluctuations of the current in the voltage-controlled regime. For
a fixed $V_g$, averaged fast-Fourier spectra, $S_I(f)$, are
obtained in the range from 0.1 to 50 Hz for different voltages $V$
across the sample, not exceeding 10 mV to satisfy linear $I-V$. It
has been established that the spectral density $S_I(f)$ is
proportional to the square of the dc current, so that the spectral
density of resistance fluctuations $S_R$ is found from the relation $%
S_R(f)/R^2\equiv S_I(f)/I^2$.

In Fig.~1 the temperature dependence of the resistance is shown
for two representative samples of different length: a) $L=1$~$\mu
$m and b) $L=$ 0.2~$\mu $m (the properties of the sample with
length 0.5~$\mu $m are very similar to those of the $1$~$\mu $m
sample). In both samples, with decreasing concentration there is a
transition to hopping transport where $R(T)$ becomes strongly
temperature dependent. At $T\lesssim 13$ K, the resistance of the
large sample is described by two-dimensional VRH: $R_{\Box
}=R_0\exp \left( T_0/T\right)
^{1/3}$, where $R_{\Box }$ is resistance 'per square' and $R_0\simeq 40$%
~kOhm.  For $T$ above $13$ K $R(T)$ is described by a simple
exponential dependence $R_{\Box }\propto \exp (\Delta E/k_BT)$,
which can suggest the NNH regime. One can see that at low
temperatures the behaviour of the small sample in Fig. 1b is quite
different: the gradient of the temperature dependence is much
smaller and it only slightly increases with decreasing
concentration.

The evolution of $R(T)$ with decreasing concentration in a
transistor structure is caused by an increase of the typical
hopping length, when the Fermi level moves down the tail of the
localised states. In the Miller-Abrahams approach the resistance
of an electron hop between sites $i$ and $j$ is determined by the
hopping distance $r_{ij\text{ }}$ and the effective activation
energy $\Delta E_{ij}$ \cite{SE}:

\begin{equation}
\rho _{ij}\simeq \rho _o\exp \left( \frac{2r_{ij}}a+\frac{\Delta E_{ij}}{k_BT%
}\right)   \label{eq1}
\end{equation}
In the VRH regime the typical hopping distance $r\simeq a\xi $, where $\xi
=\left( T_0/T\right) ^{1/3}$ , $T_0=\frac{13.5}{ga^2}$ and $g$ is the
density of states at the Fermi level. Hence,  $r$ increases with decreasing
temperature and concentration. The correlation length of the  network $%
L_c$, which is a measure of an individual link in the cluster
shown in the inset to Fig~1, also increases as $L_c\simeq r\xi
^\nu $, where $\nu =1.33$ in 2D \cite{SE}. In the NNH regime,
expected at higher temperatures when $ 2r_{ij}/a>\Delta
E_{ij}/(k_BT)$, the hopping distance is determined by the nearest
localised states and does not depend on temperature, so that the
cluster density also becomes temperature independent.

The weak temperature dependence for the small sample (Fig.~1b) can
be explained by the fact that its conducting network is reduced
into a set of parallel hopping chains \cite{RR,GM}, Fig~1b inset.
The chain conductance is larger than that of the cluster and is
less temperature dependent. The evolution of a 2D cluster into a
set of 1D chains occurs when $L_c$ becomes larger than the sample
length $L$ \cite{RR}. The hopping exponent $\xi $, estimated for
the large sample, in the range $|V_g|=1.2-1.35$~V is $\xi =2.5-5$,
which gives $ L_c\approx a\xi ^{1+\nu }=0.1-0.5$~$\mu $m for
localisation length $a\simeq a_B\simeq 100$~\AA. This satisfies
the condition $L\leq L_c$ for the majority of concentrations in
the small sample.

With this understanding of the specifics of electron hopping in
the two samples, let us turn to noise measurements to see how the
difference in the topology of the hopping paths is reflected in
$1/f$ noise. We have established that in both samples the
frequency dependence of $S_R(f)/R^2$ is close to $1/f$ in the
whole range of $V_{g\text{ }}$ and $T$. Similar to \cite{AKS}, on
top of $1/f$ noise some individual fluctuators can be revealed as
maxima in the frequency dependence of $\left[ f\times
S_R(f)/R^2\right] $, each indicating a contribution of a
Lorentzian spectrum. In the small sample the amplitude of the RTSs
is seen to be larger and can reach $\sim $10\% of the sample
resistance. (Assuming that the modulator affects only the chains
positioned nearby, the total number of chains in the small sample
is small - about 10.)

The noise power increases with decreasing concentration in the
whole temperature range in both samples. In Fig.~2 we show
examples of their normalised noise power $S_R(f)/R^2$ at frequency
$f=$1 Hz presented as a function of resistance varied by $V_g$. (A
similar experiment, although near the metal-to-insulator
transition and not in hopping, was performed on In$_x$O$_{1-x}$
films in \cite{Cohen}, where noise power showed saturation with
increasing resistance.) In the small sample the growth is weaker
at large resistances, where the formation of hopping chains is
expected.

Comparison of the temperature dependences of noise and resistance
in the large sample in the whole temperature range shows an
interesting tendency, Fig. 3. With increasing $T$ and approaching
the transition from the VRH to the activation regime, the
dependence $R(T)$ becomes stronger, while the temperature
dependence of $S_R(f)/R^2(T)$ weakens. (The short sample has shown
a similar tendency, but with a weaker temperature dependence of
noise at lower temperatures.)

We can explain these results within the modulation approach, which
we modify here for the case of 2D hopping. According to model
\cite{Kozub}, the noise power is determined by two factors: a) the
number of links in the cluster $N$, i.e. the cluster density; b)
the modulation intensity of the critical (dominant) hop in the
link. To find the integral noise, random fluctuations $(\delta
\rho _i,\delta \rho _i)_\omega $ of the critical resistances $\rho
_i$ are averaged over $N$ links of the cluster, to give
\begin{equation}
\frac{S_R(f)}{R^2}\sim \frac 1{N^2}\Sigma _i\frac{(\delta \rho _i,\delta
\rho _i)_\omega }{\rho _i^2}\sim \frac 1N\frac{(\delta \rho _i)_\omega ^2}{%
\rho _i^2}  \label{eq2}
\end{equation}
The modulation of resistance $\rho _i$ is caused by the shift
$\delta V$ of the energy $\Delta E_{ij}$ in the critical hop,
Eq.~\ref{eq1}, so that $\mid \delta \rho _i/\rho _i\mid $ $\sim
\min (\delta V/k_BT$, $1)$. (The maximum
possible modulation of $\sim $100 \% corresponds to large energy change, $%
\delta V/k_BT>1$, when the critical path itself is altered.) The
increase of noise in Fig. 2 with decreasing concentration can then
be viewed as being primarily due to a decrease in the number of
fluctuating elements $N=\left( L/L_c\right) ^2$, when, with
increasing hopping length $r$, the cluster becomes less dense.
Slow increase of noise in the small sample indicates that the
number of statistically chosen, most conductive, chains does not
vary significantly when electron concentration is decreased
\cite{RR}.

The temperature effect on noise in Fig. 3 can also be understood
in terms of its effect on the critical cluster density. In VRH,
the number of links $N$ rapidly increases when the temperature
goes up, and this is why the noise drops. However, in the
activation regime, the cluster size no longer increases and the
noise dependence on $T$ weakens.

To analyse the temperature dependence of the noise power quantitatively, one
has also to take into account the temperature dependence of the modulation
intensity $M(T)$. Introducing the modulator density $P(E,\tau )$, where $E$
is the energy of the modulator level and $\tau $ is the relaxation time, and
assuming that $P(\tau )$ to be exponentially broad to produce the $1/f$
spectrum, one can write in 2D :
\begin{eqnarray}
M &=&\frac{(\delta \rho _i)_\omega ^2}{\rho _i^2}\sim
\frac{P(T)T}\omega \smallint \min \left( \frac{\left( \delta
V(D)\right) ^2}{T^2},1\right)d^2R
\label{eq3} \\
\  &\sim &\frac{P(T)T}\omega D_T^2(T)  \nonumber
\end{eqnarray}
In this expression $\delta V(D,r)$ is the potential variation on the length $%
r$ of the critical hop, $\delta V\sim r(\partial V/\partial D)$,
which is caused by an electron transition between the two states
of the modulator positioned at distance $D$ from the critical hop,
inset to Fig. 2a. The factor $T$ in this expression describes the
number of active modulators whose occupancy fluctuates at a given
temperature. The area $D_T^2(T)$ gives the number of the strongest
($\sim $100\%) modulators that are able to alter the critical hop.
Such modulators have to be positioned within the distance $D_T$ to
the hop, such that $\delta V(D_T)\simeq k_BT$. The distance
$D_T(T)$ depends on the details of the potential $V(D)$ produced
by the modulator. Namely, for a dipole potential, when the
separation $l$ of the two states of the modulator is smaller than
the distance $D$ from the modulator to the critical hop, $ \delta
V\sim r(\partial V/\partial D)\propto rl/RD^{3}$ and $D_T^3\sim
e^2rl(\omega )/(\chi k_BT)$. Then the efficiency $M(T)$ increases
with temperature in VRH as $T\left( r/T\right) ^{2/3}\propto
T^{1/9}$ (taking into account that $r\sim a\xi $ and $\xi =$
$\left( T_0/T\right) ^{1/3}$). In the opposite case of large $l$,
the potential of the modulator is determined by the Coulomb
potential of the state closest to the critical hop, so that
$\delta V\propto r/D^2$ and $D_T^2\sim e^2r/(\chi k_BT)$. In this
case the modulation efficiency $M(T)\propto T^{-1/3}$ decreases
with increasing temperature.

To get information about the temperature dependence $M(T)$ in VRH
of the large sample, we have performed noise measurements at the
condition when, at different temperatures, the resistance of the
sample is kept constant by adjusting $V_g$. For VRH this
corresponds to the same density of the cluster, so that the
temperature dependence of noise is only due to the variation of
the modulation efficiency. The results are shown in Fig. 4. The
increase of noise with decreasing temperature at $T<$13 K{\bf \
}is an indication that the modulator potential is not of a dipole
type but produced by a Coulomb potential of a single state.

One can see that with increasing $T$ and transition to the
exponential regime, the temperature dependence of noise at a fixed
sample resistance changes its sign, Fig. 4. This results from the
fact that in this regime the noise measured at a fixed $V_g$ only
weakly decreases with increasing $T$, while the resistance drops
significantly. In addition, the noise strongly depends on
resistance, Fig. 2a. Therefore, when $T$ is increased and $V_g$ is
then adjusted to keep the resistance constant, the measured noise
increases. In this regime the cluster becomes temperature
independent, and the noise can be caused by both the modulation
\cite{Kozub} and fluctuation of concentration \cite{KoShk}. There
is a difficulty, however, in analysing this regime quantitatively,
as in our case it is not conventional NNH. This is seen in Fig. 1a
where the activation energy of $R(T)$ increases with decreasing
concentration, which means that not all the states of the impurity
band are involved in hopping and their number depends on the Fermi
level position. This can be due to the fact that at high $T$ one
should take into account the energy dependence of the density of
states in the tail of the impurity band, which modifies the
character of hopping \cite {Shapiro}.

Let us come back to the low-temperature VRH where our results show
the domination of the modulation source of noise. We can now
combine the information about the modulation efficiency $M(T)$ in
the VRH regime with the number of links $N(T)=\left( L/r\xi ^\nu
\right) ^2$, and get the temperature dependence of the noise power
in VRH of the large sample:

\begin{eqnarray}
\frac{S_R(f)}{R^2} &\sim &\frac 1{N(T)}M(T)\sim \left( \frac{r\xi ^\nu }L%
\right) ^2\frac{P_0T}\omega \left( \frac{e^2r}{\chi k_BT}\right)
\label{eq4} \\
&\propto &\xi ^{2\nu +3}\propto T^{-1.89}  \nonumber
\end{eqnarray}
This result is in satisfactory agreement with $S_R(T)/R^2$
presented in Fig.~3, within the experimental error in the measured
noise power. Using Eq.~\ref{eq4} and relation $\xi =\ln
(R(T)/R_0)$ we can also obtain a relationship between the noise
power and the resistance: $S_R(f)/R^2\propto \xi ^{2\nu +3}=\ln
^{5.66}(R_{\Box }/R_0)$. Comparison of this dependence with data
in Fig.~2a also shows reasonable agreement.

In conclusion, we have demonstrated that the site-energy
modulation model accounts for the properties of $1/f$ noise in the
VRH regime of 2D hopping in an {\em n}-GaAs channel. We have been
able to separate different components of $1/f$ modulation noise
that are responsible for its temperature and resistance
dependence. It has also been shown that the noise measurements are
sensitive to the transition between different hopping regimes, as
well as to the difference in the topology of the hopping cluster
in large and small samples. \\

We are grateful to S.M.Kogan, M.E.Raikh and B.I.Shklovskii for
helpful discussions, and the Royal Society, EPSRC, the RFBR (grant
99-02-17387), and MNTP ``Physics of Solid State Nanostructures''
(grant 97-1052) for financial support.

\begin{figure}[tbp]
\caption{Temperature dependence of resistance at different gate
voltages for, from bottom to top: a) long sample, $V_g=-0.7$,
$-1.24$, $-1.34$, $ -1.39 $, $-1.45$, $-1.52$~V, and b) short
sample $V_g=-0.22$, -0.78, -1.25, -1.38, -1.46, -1.52, -1.58~V.
Dashed lines in a) show agreement with 2D variable-range hopping.
The insets illustrate the structure of the hopping paths - a
cluster in the large sample and chains in the small sample. Arrow
$I$ shows the direction of the current.} \label{Fig1}
\end{figure}

\begin{figure}[tbp]
\caption{Dependence of the noise power (at $f=1$~Hz) on resistance
varied by $V_g$: a) long sample, b) short sample. The broken lines
show fits by the relation $S_R(f)/R^2\propto \ln ^{5.66}(R/R_0)$,
with $ R_0=40$~k$\Omega$. Inset: a diagram of a two-site modulator
acting on the critical hop;  $r$ is the length of the critical hop
with resistance $\rho_{ij}$ , $l$ is the separation of the two
states of the modulator, $D$ is its distance from the critical
hop. } \label{Fig2}
\end{figure}

\begin{figure}[tbp]
\caption{Temperature dependence of resistance $R_{\Box }$ (open
circles) for the long sample, with dashed lines $R(T)\propto
exp(T_0/T)^{1/3}$, and its normalised noise power (closed circles)
at $f=1$~Hz, where the dashed lines through the points are guides
to the eye. The bold dashed line is plotted using Eq.~\ref{eq4}.}
\label{Fig3}
\end{figure}

\begin{figure}[tbp]
\caption{Temperature dependence of noise at $f=1$~Hz when the
resistance is kept constant by adjusting $V_g$. Different symbols
correspond to different values of $R$ indicated in the plot. The
dotted curve shows the dependence $M(T)\propto T^{-1/3}$.}
\label{Fig4}
\end{figure}

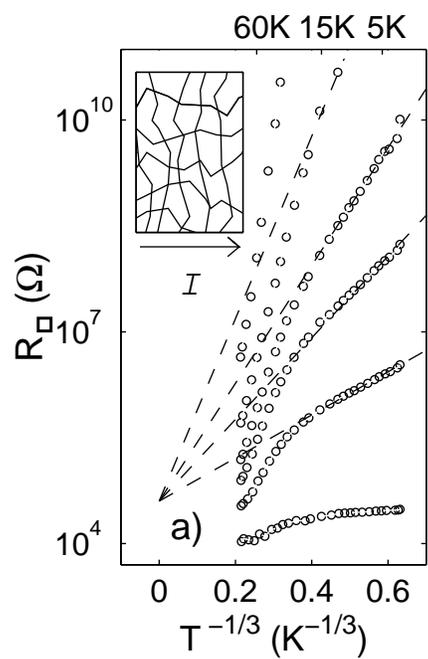
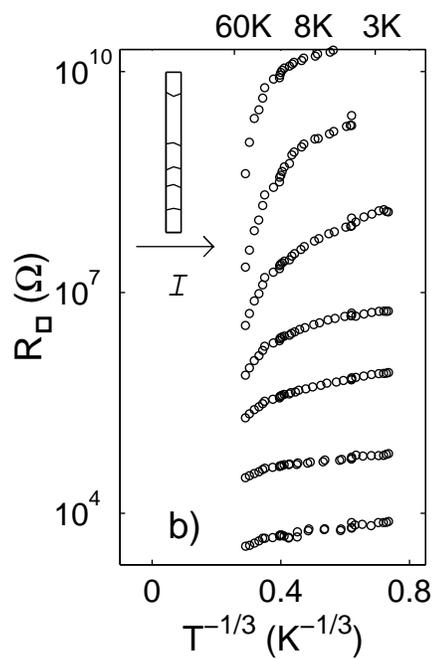
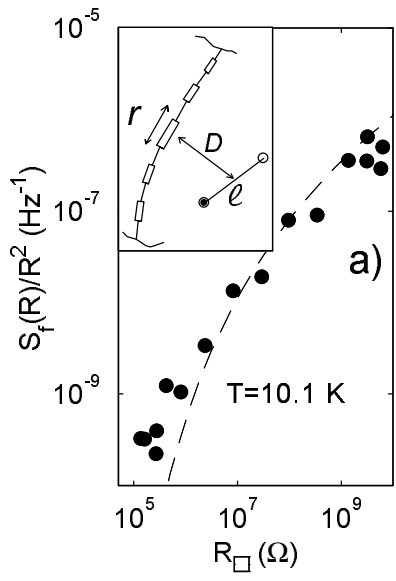
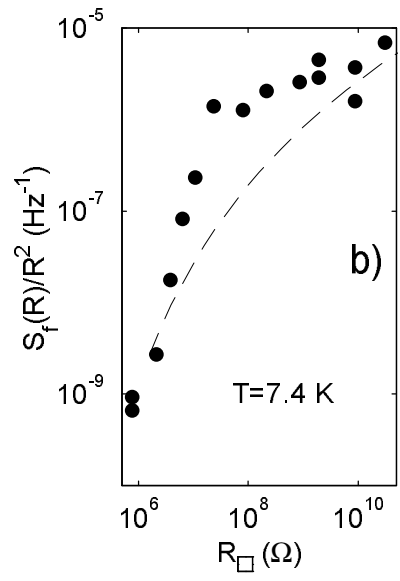
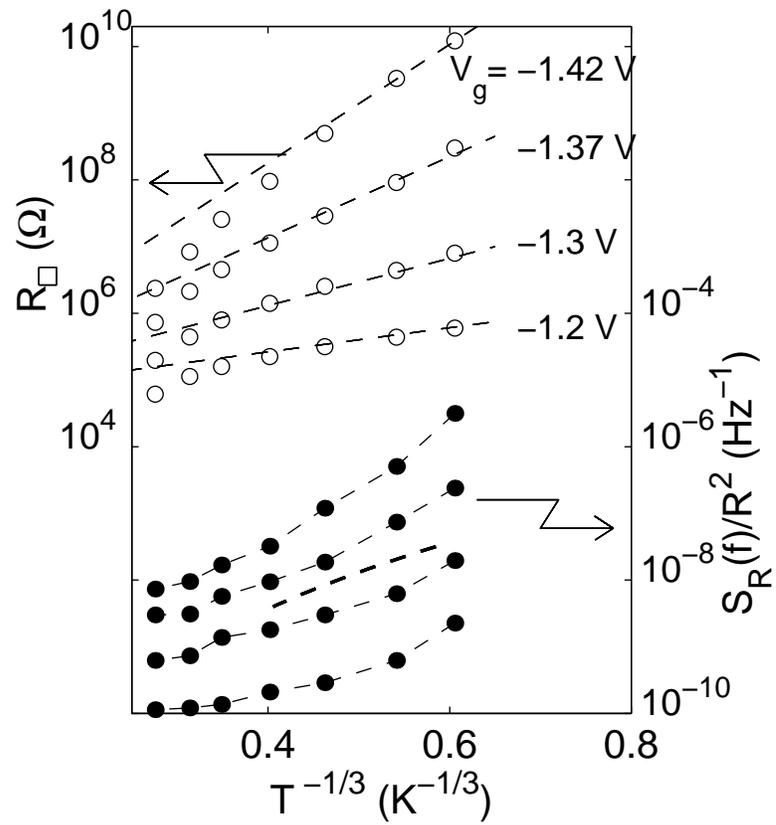
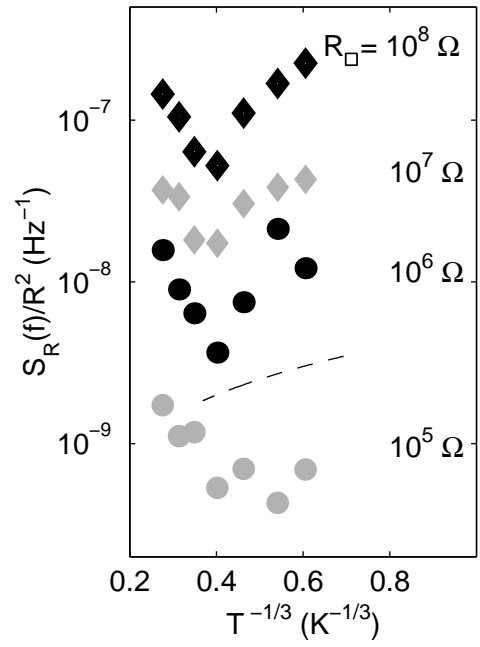
\end{document}